\documentclass[aps,pre,twocolumn,groupedaddress,showpacs,longbibliography]{revtex4-1}
\pdfoutput=1
\usepackage{amssymb,amsmath,xcolor,graphicx}
\begin{document}

%\preprint{}

\title{Approaching Carnot efficiency at maximum power in linear response regime}

\author{Marcus V. S. Bonan\c{c}a}
\email[]{mbonanca@ifi.unicamp.br}
%\thanks{}
\affiliation{Instituto de F\'isica `Gleb Wataghin', Universidade Estadual de Campinas, 13083-859, Campinas, S\~ao Paulo, Brazil}


\date{\today}

\begin{abstract}
We construct an example of heat engine whose efficiency at maximum power breaks down the previously derived bounds in the linear response regime. Such example takes a classical harmonic oscillator as the working substance undergoing a finite-time Otto cycle. Using a specific kind of shortcut to adiabaticity, valid only in the linear response regime, quasistatic work is performed at arbitrarily short times. The cycle duration is then reduced to the sum of relaxation times during the thermalization strokes exclusively. Thus, power is maximum since the work is maximum (quasistatic work) and the cycle duration is minimum. Efficiency at maximum power can be made arbitrarily close to Carnot efficiency with an appropriate choice of the ratio between the temperatures of the two heat baths. 
\end{abstract}

%\pacs{05.45.Mt}
%\keywords{relaxation, fluctuation-dissipation theorem, linear response}

\maketitle


%%%%%%%%%%%%%%%%%%%%%%%%%%%%%%%%%%%%%%%%%%%%%%%%%%%%%%%%%%%%%%%%%%%%%%%%
\section{Introduction}
%\paragraph{Introduction.}

The issue of maximum efficiency of heat engines is considered the foundational problem of classical thermodynamics. Due to the technological advances of the last decades, it has become possible to investigate such problem using microscopic heat engines \cite{klimovsky2018,benenti2017,rossnagel2016,martinez2016,pekola2015,rossnagel2014,brantut2013,blickle2012,abah2012,steeneken2011,hugel2002} and test the thermodynamic principles in this new context. On the theoretical side, different lines of research have developed out of this revisited problem of thermodynamic efficiency. For instance, it is well-known that the Carnot or reversible heat engine produces no power. It is natural to ask then whether the efficiency of finite-time heat engines under the constraint of maximum power also follows some kind of universal bound \cite{broeck2005,rezek2006,schmiedl2008,izumida2008,tu2008,izumida2009,esposito2009,esposito2010,benenti2011,broeck2012,broeck2013,
guo2013,seifert2013,seifert2014,uzdin2014,jiang2014,lutz2017}. These investigations naturally have branched into several important questions about the trade-off between power and efficiency \cite{gaveau2010,seifert2015,broeck2015,broeck2016a,raz2016,tasaki2016,broeck2016b,holubec2016,seifert2018,ponmurugan2018a,ponmurugan2018b,deffner2018} that touch the specific problem of having Carnot efficiency with finite power \cite{allah2013,gadas2015,campisi2016a,campisi2016b,johnson2018}. Research on this issue in particular has also benefited from the study of efficiency fluctuations in small-scale heat engines using advanced methods in stochastic thermodynamics \cite{broeck2014,gingrich2014,verley2014,polettini2015,polettini2017}. Finally, the possible effects due to shortcuts to adiabaticity on power and efficiency of quantum heat engines have added additional perspectives to the well-established investigations mentioned above \cite{campo2013,jarzynski2013,gong2013,tu2014,campo2014,campo2017,kosloff2017,abah2017}.

It has been argued that the efficiency at maximum power does follow universal bounds at least in the linear response regime \cite{esposito2009,esposito2010,broeck2013,guo2013}. In this approach, linear response is understood as a regime of small entropy production in which thermodynamic fluxes can be linearly expressed in terms of the corresponding thermodynamic forces. These linear relations are formulated in terms of the well-known Onsager coefficients whose properties and physical consequences have been extensively discussed in linear irreversible thermodynamics \cite{groot1984,seifert2015,broeck2015}. However, these linear relations neglect the possible delay in the response of the system due to the disturbance generated by the thermodynamic forces since fluxes $J_{k}$ and forces $X_{k}$ are evaluated \emph{at the same instant of time}. Such delay can be taken into account through the following linear relation
\begin{equation}
J_{k}(t) = \int_{-\infty}^{t}ds\,\sum_{l}\Phi_{k l}(t-s)X_{l}(s)\,,
\label{eq:irrethermo}
\end{equation}
where $\Phi_{k l}(t)$ describes the response of the system to impulsive forces. Linear response regime should be understood then as the class of close-to-equilibrium thermodynamic processes in which \emph{the most general} linear relation between fluxes and forces take place \cite{kubo1985}. This means that both fast and slow processes are allowed as long as fluxes are linearly related to forces. Thus, the equations of linear irreversible thermodynamics can be recovered from (\ref{eq:irrethermo}) when the process is sufficiently slow compared to the tendency of the system of interest to go back to equilibrium. In this case, the delay is negligible and fluxes respond almost instantaneously to forces.

The difference between results obtained from these two kinds of linear processes, namely, fast and slow, can be illustrated using for instance the irreversible or excess work, denoted here by $W_{ex}$. This quantity is defined as the thermodynamic work $W$ that has been performed along a given process minus its corresponding quasistatic value $W_{qs}$ \cite{sivak2012,bonanca2014,acconcia2015a,bonanca2018}. For fast processes, it has been shown that there can be shortcuts to adiabaticity in which the quasistatic value of work can be achieved in finite time with zero additional cost when the system is thermally isolated \cite{acconcia2015b}. Consequently, these processes allow us to perform a finite-time Otto cycle in which the total work is equal to the corresponding value in the quasistatic cycle and hence maximum. Additionally, the time spent in the strokes in which the system is thermally isolated can be arbitrarily small due to the shortcuts \cite{acconcia2015b}. Thus, the cycle duration is minimized since it is reduced to the sum of time intervals along the thermalization strokes.

In what follows we describe more carefully the details of our finite-time heat engine. However, we emphasize that the use of shortcuts to adiabaticity is essential in our proposal. The shortcuts we consider are possible only in processes described by Eq.~(\ref{eq:irrethermo}) which were not considered in previous analysis of efficiency at maximum power in the linear response regime.

%%%%%%%%%%%%%%%%%%%%%%%%%%%%%%%%%%%%%%%%%%%%%%%%%%%%%%%%%%%%%%%%%%%%%%%%
%\section{Finite-time Otto cycle using linear response shortcuts \label{sec:lrotto}}
%\paragraph{Linear response shortcuts.}
\section{Linear response shortcuts}

We start constructing an Otto cycle using as a working substance the following classical harmonic oscillator,
\begin{equation}
H = \frac{p^{2}}{2m} + \lambda\frac{q^{2}}{2}\,.
\label{eq:harmosc}
\end{equation}
The spring constant $\lambda$ will play the role of the externally controlled parameter through which we can perform work when the oscillator is thermally isolated. We also need two heat baths at different temperatures $T_{1}$ and $T_{2}$ for the strokes responsible for heat exchange. 

\begin{figure}
\includegraphics[width=.42\textwidth]{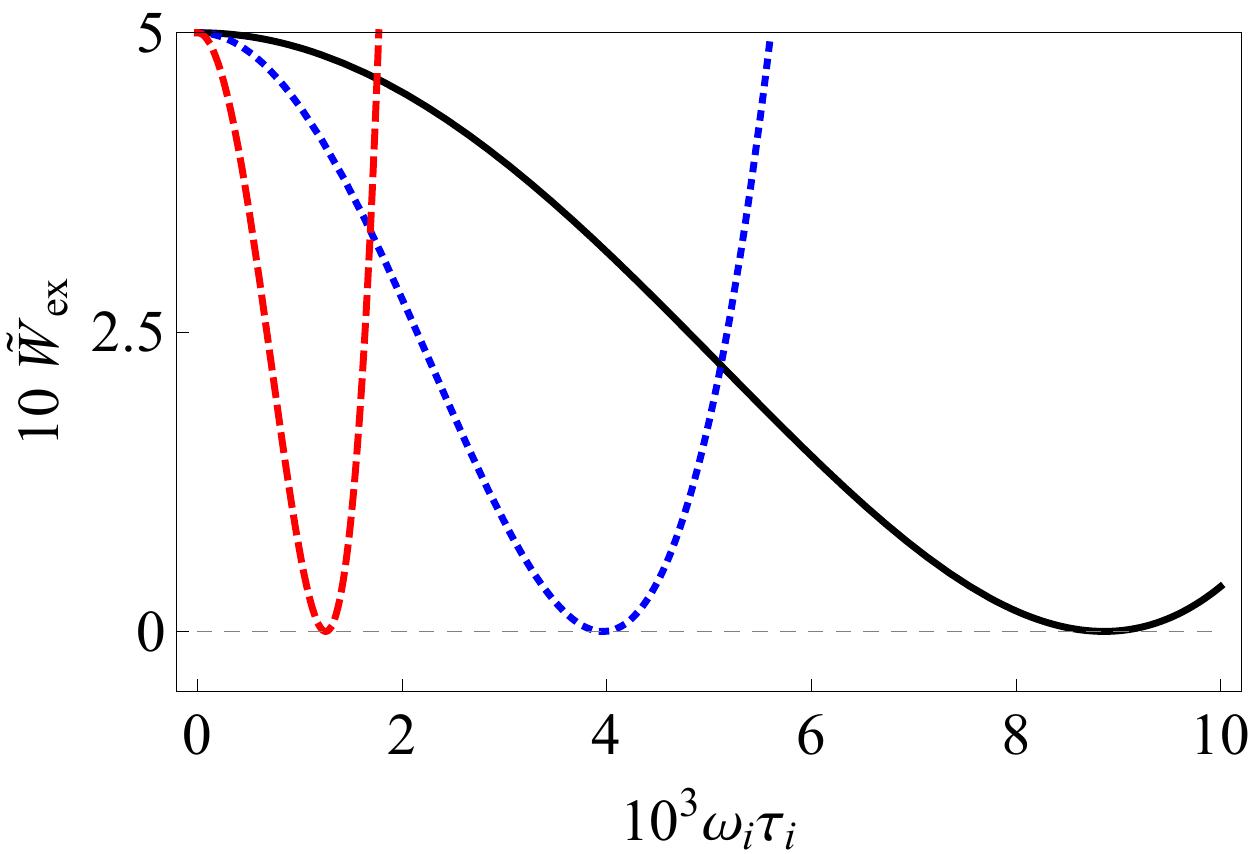}
\caption{\label{fig:wexshort} (color online) Short time behavior of the excess work $W_{ex}$ given by Eq.~(\ref{eq:wexlr}) using the relaxation function (\ref{eq:relaxf}) and the protocols (\ref{eq:protocg}) with the following values of $a_{i}$: $2\times 10^{4}$ (black solid line), $10^{5}$ (blue dotted line) and $10^{6}$ (red dashed line).}
\end{figure}

The oscillator starts in thermal equilibrium (described by a Boltzmann-Gibbs distribution) with the heat bath at temperature $T_{1}$ and $\lambda=\lambda_{1}$. Once the thermal contact is broken, we vary the parameter $\lambda$ from $\lambda_{1}$ to $\lambda_{2}=\lambda_{1}+\delta\lambda$ according to the following protocol, $\lambda_{1}(t)=\lambda_{1} + \delta\lambda\,g_{i=1}(t)$, where
\begin{equation}
g_{i}(t) = \frac{t-t_{0,i}}{\tau_{i}} + a_{i}\sin{\left[ 2\pi\frac{(t-t_{0,i})}{\tau_{i}} \right]}\,.
\label{eq:protocg}
\end{equation}
We denote the initial time by $t_{0,i}$ and the function $g_{i}(t)$ is such that $g_{i}(t_{0,i})=0$ and $g_{i}(t_{0,i}+\tau_{i})=1$, which means that $\tau_{i}$ is the duration of the protocol or switching time. The reason behind this choice is the following: it was shown in Refs.~\cite{acconcia2015a,acconcia2015b} that this protocol leads to a kind of shortcut to adiabaticity \cite{campo2013,jarzynski2013} (adiabaticity in the mechanical sense) in the linear response regime. In other words, it is possible to choose values of $\tau_{1}$ and $a_{1}$ such that the thermodynamic work performed along the finite-time process is equal to the value $W^{qs}_{1}$ obtained after performing the corresponding quasistatic process. Moreover, this is achieved without adding extra terms (the so-called counterdiabatic terms \cite{campo2013}) to the Hamiltonian (\ref{eq:harmosc}). This means that we can drive the system in finite time having $W^{qs}_{1}$ as the only energetic cost for it. Additionally, it was also shown in Ref.~\cite{acconcia2015a} that, by choosing $a_{1}$ properly, the values of $\tau_{1}$ for which the thermodynamic work is equal to $W^{qs}_{1}$ can be arbitrarily close to zero (see Fig.~\ref{fig:wexshort} and Eq.~(\ref{eq:lrtau})). For a numerical confirmation of this prediction see Ref.~\cite{acconcia2015a}.

Shortcuts carried by (\ref{eq:protocg}) are restricted to systems such as the harmonic oscillator and the spin-1/2 in a magnetic field when thermally isolated \cite{acconcia2015b}. For anharmonic oscillators, shortcuts due to (\ref{eq:protocg}) were only approximately verified \cite{acconcia2015a}. Protocol (\ref{eq:protocg}) is likely to lose its effectiveness when the spring constant of the harmonic potential is changed in the presence of anharmonic terms.

The above-mentioned claims follow from the expression below for the excess work $W_{ex} = W - W_{qs}$ in linear response theory \cite{acconcia2015a}, 
\begin{eqnarray}
W_{ex}= \frac{(\delta\lambda)^{2}}{2}\int_{t_{0,i}}^{t_{0,i}+\tau_{i}}dt\int_{t_{0,i}}^{t_{0,i}+\tau_{i}}dt'\,\Psi(t-t')\,\dot{g}_{i}(t)\,\dot{g}_{i}(t')\,,\nonumber\\
\label{eq:wexlr}
\end{eqnarray}
where $\dot{g}_{i}(s)$ and $\dot{g}_{i}(s')$ denote the derivatives with respect to $t$ and $t'$. We denote by $\Psi(t)$ the so-called \emph{relaxation function} \cite{kubo1985} which, in our case, must be calculated for a thermally isolated harmonic oscillator. In this case, $\Psi(t)$ is basically the equilibrium auto-correlation function of the observable $\partial H/\partial\lambda = q^{2}/2$ and reads \cite{acconcia2015a}
\begin{equation}
\Psi(t) = \frac{\cos{(2\omega_{i} t)}}{4\beta_{i}\lambda^{2}_{i}} = \Psi(0)\cos{(2\omega_{i} t)}\,,
\label{eq:relaxf}
\end{equation}
where $\omega_{i}=\sqrt{\lambda_{i}/m}$ is the natural frequency of oscillation for a given $\lambda=\lambda_{i}$ and $\beta_{i}\equiv(k_{B}T_{i})^{-1}$ with $i=1,2$. The derivation of Eq.~(\ref{eq:wexlr}) can be briefly sketch as follows: the work $W$ is expressed as
\begin{equation}
W = \int_{t_{0,i}}^{t_{0,i}+\tau_{i}}dt\,\frac{d\lambda_{i}}{dt} \overline{\frac{\partial H}{\partial\lambda_{i}}}(t) = 
\delta\lambda \int_{t_{0,i}}^{t_{0,i}+\tau_{i}}dt\,\dot{g}_{i}(t) \overline{\frac{\partial H}{\partial\lambda_{i}}}(t)\,,
\label{eq:work}
\end{equation}
where $\overline{A}$ denotes the out-of-equilibrium average of the observable $A$ (the initial equilibrium distribution is perturbed due to driving and therefore $W$ is obtained in terms of the out-of-equilibrium average of $\partial H/\partial\lambda$). Linear response theory yields \cite{kubo1985}
\begin{equation}
\overline{\frac{\partial H}{\partial\lambda_{i}}}(t) = \left\langle \frac{\partial H}{\partial\lambda_{i}}\right\rangle +\chi \delta\lambda g_{i}(t) +\delta\lambda\int_{t_{0,i}}^{t}dt'\,\Psi(t-t')\,\dot{g}_{i}(t')\,,
\label{eq:response}
\end{equation}
where $\langle\cdot\rangle$ denotes an average over the initial canonical ensemble, $\chi$ takes into account the possibility of an instantaneous response and the last term describes the delayed response in terms of $\Psi(t)$. Equation (\ref{eq:response}) resembles the linear relation (\ref{eq:irrethermo}) between fluxes and forces. Plugging (\ref{eq:response}) into (\ref{eq:work}) leads to Eq.~(\ref{eq:wexlr}) after some algebra (for more details see Ref.~\cite{acconcia2015a}).

The excess work is defined as the difference between the thermodynamic work $W$ along a given process and its corresponding value $W_{qs}$ in the quasistatic regime. Hence every time $W_{ex}$ is zero for a finite switching time, we have found a specific protocol for which $W = W_{qs}$ in finite time. Figure~\ref{fig:wexshort} shows that this is achieved using protocols (\ref{eq:protocg}) for specific values of $\tau_{i}$ given by (see Appendix~\ref{sec:app} for a short derivation and Ref.~\cite{acconcia2015a} for more details)
\begin{equation}
\omega_{i}\tau_{i} = \pi/(1+2\pi a_{i})^{1/2}\,.
\label{eq:lrtau}
\end{equation}
In addition, these values can be made arbitrarily small by choosing $a_{i}$ properly. This is going to be crucial in the analysis of the power generated by our engine.

The first stroke of our Otto cycle is hence a finite-time switching of $\lambda_{1}$ to $\lambda_{2}$ keeping the working substance thermally isolated. The second stroke is simply a thermalization process. We bring our oscillator in contact to the second heat bath at temperature $T_{2}$ and wait until it thermalizes while keeping $\lambda=\lambda_{2}$. Thus, we assume that this process lasts a finite time interval equal to the relaxation time $\tau^{R}_{2}$ after which the system is in a Boltzmann-Gibbs distribution (relaxation times of the order of microseconds have been measured in ion traps \cite{rossnagel2016}). Before the third stroke starts, we break the thermal contact between the second heat bath and the oscillator. After that, we drive the system in a finite time interval $\tau_{2}$ switching $\lambda$ from $\lambda_{2}$ to $\lambda_{1}$ while keeping the oscillator thermally isolated. This is done using the protocol $\lambda_{2}(t)=\lambda_{2} - \delta\lambda\,g_{2}(t)$, with $g_{2}(t)$ given by Eq.~(\ref{eq:protocg}) with $i=2$. As in the first stroke, the values of $\tau_{2}$ and $a_{2}$ are chosen in such way that the thermodynamic work is equal to the corresponding quasistatic value $W^{qs}_{2}$. The final stroke consists of bringing the oscillator back in contact to the heat bath at temperature $T_{1}$ and waiting for the thermalization process whose duration we denote by $\tau^{R}_{1}$ and after which we assume the system is back to the initial canonical distribution.

We model each thermalization stroke as an underdamped Brownian motion in a harmonic trap. Hence it is well known for this model that initial nonequilibrium distributions approach the Boltzmann-Gibbs distribution exponentially fast. The expression we take in the following for the $\tau^{R}_{1,2}$ also relies on the above-mentioned model of Brownian motion.

%%%%%%%%%%%%%%%%%%%%%%%%%%%%%%%%%%%%%%%%%%%%%%%%%%%%%%%%%%%%%%%%%%%%%%%%%
%\section{The finite-time Otto engine \label{sec:ftotto}}
\section{Finite-time Otto engine}

We shall calculate now the exchange of mechanical and thermal energy along each of the previously described strokes of our finite-time Otto cycle. During the first stroke, the oscillator is thermally isolated and driven using the shortcut described in the previous section. Thus, the heat exchanged is zero, the excess work is also zero and the work $W_{1}$ performed after a time interval $\tau_{1}$ is 
\begin{equation}
W_{1} = W_{1}^{qs} = k_{B}T_{1}\left[ \left( \frac{\lambda_{2}}{\lambda_{1}}\right)^{1/2} - 1\right]\,.
\end{equation}

This expression is obtained from the invariance of action along a quasistatic change of $\lambda$ and the equipartition theorem. In other words, when the spring constant of a harmonic oscillator is changed quasistatically, its initial energy $E_{1}$ is related to its final energy $E_{2}$ through the relation $E_{2}/E_{1} = (\lambda_{2}/\lambda_{1})^{1/2}$.

The same reasoning is valid for the third stroke since we vary $\lambda$ from $\lambda_{2}$ to $\lambda_{1}$ using another shortcut while the oscillator is again kept thermally isolated. Thus, the work $W_{2}$ performed after a time interval $\tau_{2}$ reads
\begin{equation}
W_{2} = W_{2}^{qs} = k_{B}T_{2}\left[ \left( \frac{\lambda_{1}}{\lambda_{2}}\right)^{1/2} - 1\right]\,.
\end{equation}
It is worth emphasizing at this point that both $\tau_{1}$ and $\tau_{2}$ obey Eq.~(\ref{eq:lrtau}) for given $a_{1}$ and $a_{2}$.

The work performed along strokes 2 and 4 is zero simply because $\lambda$ is held fixed. Hence we just have to care about the exchanged heat between the oscillator and the heat bath in both strokes. They can be obtained as follows: after the first stroke, the internal energy of the oscillator is the sum of the internal energy $\langle E\rangle_{1}=k_{B}T_{1}$ it had before this stroke started and the mechanical energy $W_{1}$ transferred at the end of the process. As the internal energy after the second stroke is $\langle E\rangle_{2}=k_{B}T_{2}$, the heat exchanged with the second heat bath is easily obtained from the First Law and reads
\begin{eqnarray}
Q_{2} &=& \langle E\rangle_{2} - \langle E\rangle_{1}\left( \frac{\lambda_{2}}{\lambda_{1}}\right)^{1/2} \nonumber\\
&=& k_{B}T_{2} \left[ 1 - \frac{T_{1}}{T_{2}}\left( \frac{\lambda_{2}}{\lambda_{1}}\right)^{1/2}\right]\,,
\end{eqnarray}
where we have used the equipartition theorem to express the internal energies in terms of the temperatures. Analogously, the heat exchanged between the oscillator and heat bath along the fourth stroke reads
\begin{equation}
Q_{1} = k_{B}T_{1} \left[ 1 - \frac{T_{2}}{T_{1}}\left( \frac{\lambda_{1}}{\lambda_{2}}\right)^{1/2}\right]\,.
\end{equation}

We are now ready to compute the work performed along the cycle. It is given by
\begin{eqnarray}
W_{cycle} &=& W_{1} + W_{2} \nonumber \\
&=& -k_{B}T_{2} \left[ 1-\left( \frac{\lambda_{1}}{\lambda_{2}}\right)^{1/2}\right] \left[1-\frac{T_{1}}{T_{2}}\left( \frac{\lambda_{2}}{\lambda_{1}}\right)^{1/2} \right]. \nonumber\\
\label{eq:wcycle}
\end{eqnarray}
The sign convention we are using here is such that negative work means work performed by the oscillator. If we restrict our cycle to $\lambda_{1} < \lambda_{2}$, expression (\ref{eq:wcycle}) is negative only if
\begin{equation}
\frac{T_{1}}{T_{2}} < \left( \frac{\lambda_{1}}{\lambda_{2}}\right)^{1/2}\,.
\label{eq:enginecond}
\end{equation}
This condition implies that $Q_{1} < 0$ and $Q_{2} > 0$, which means, in our sign convention, that the oscillator absorbs heat from the heat bath at temperature $T_{2}$ and releases heat into the heat bath at $T_{1}$.

%%%%%%%%%%%%%%%%%%%%%%%%%%%%%%%%%%%%%%%%%%%%%%%%%%%%%%%%%%%%%%%%%%%%%%%%
%\section{Efficiency and power \label{sec:effpow}}
\section{Efficiency and power}

The efficiency $\eta_{LR}$ of our engine can be straightforwardly obtained from the previous results. It reads
\begin{equation}
\eta_{LR} = \frac{|W_{cycle}|}{Q_{2}} = 1-\left( \frac{\lambda_{1}}{\lambda_{2}}\right)^{1/2}\,,
\label{eq:effic1}
\end{equation}
where $(\lambda_{1}/\lambda_{2})^{1/2}$ plays the role of compression ratio \cite{kosloff2017}. Due to the inequality (\ref{eq:enginecond}), $\eta_{LR}$ is certainly below the Carnot efficiency, $\eta_{C}=1-(T_{1}/T_{2})$, of an engine that would operate exclusively between heat baths 1 and 2 in the quasistatic regime. However, $\eta_{LR}$ can be arbitrarily close to $\eta_{C}$ as long as condition (\ref{eq:enginecond}) is fulfilled. For instance, given $\lambda_{1}<\lambda_{2}$, we might choose $T_{1}/T_{2}$ as
\begin{equation}
 \left( \frac{T_{1}}{T_{2}} \right)^{\frac{1}{\alpha}}=\left( \frac{\lambda_{1}}{\lambda_{2}} \right)^{1/2}\,,
\label{eq:choice}
\end{equation}
with $\alpha > 1$. This choice certainly allows for an efficiency greater than Curzon-Ahlborn \cite{yvon1955,novikov1958,curzon1975,broeck2005} for $1<\alpha<2$, but also implies that, for small $\eta_{C}$,
\begin{equation}
\eta_{LR} = \frac{\eta_C}{\alpha} +O(2)\,.
\label{eq:effic2}
\end{equation}
showing that $\eta_{LR}$ can be arbitrarily close to $\eta_C$ as $\alpha$ approaches 1. Equations (\ref{eq:effic2}) and (\ref{eq:power}) are the main results of this paper since they show that our finite-time heat engine breaks down the universality of efficiency at maximum power in the linear response regime \cite{esposito2009,esposito2010,broeck2013,guo2013}. We have already shown that the protocol $\lambda(t)$ gives the maximum value of $W_{cycle}$ in the linear response regime, i.e., when $\delta\lambda/\lambda_{1}\ll 1$. Due to (\ref{eq:choice}), this implies that
\begin{equation}
\frac{\delta T}{T_{1}} = \left(1+\frac{\delta\lambda}{\lambda_{1}}\right)^{\alpha/2} = \frac{\alpha}{2}\frac{\delta\lambda}{\lambda_{1}}+O(2)\,,
\end{equation}
where $\delta T\equiv T_{2}-T_{1}$. Thus, the heat exchange along strokes 2 and 4 also occur in the linear response regime.

To calculate the maximum power, we first need to calculate the time interval $\tau_{cycle}$ necessary to complete the finite-time cycle. This is obtained by summing up the duration of each stroke. As we discussed previously, $\tau_{1,2}$ can be made identical and arbitrarily close to zero by choosing $a_{1,2}$ appropriately. Concerning the relaxation times $\tau_{1,2}^{R}$, it is important to stress that our working medium is a Brownian particle that has no intrinsic relaxation time when disconnect from a heat bath. Hence, we take the $\tau_{1,2}^{R}$ as the time intervals necessary to complete the thermalization processes between the particle and the heat baths. This contrasts with previous analysis of systems in the weak-dissipation limit. Assuming an \emph{underdamped} regime, the $\tau_{1,2}^{R}$ are proportional to the period of oscillation, i.e., $\tau_{1,2}^{R} = \kappa/\omega_{1,2}$ (see, for example, Sec.~IV of Ref.~\cite{bonanca2014}). Choosing $\omega_{1}\tau_{1}=\omega_{2}\tau_{2}=\epsilon$, with $\epsilon\ll 1$, the duration of the cycle $\tau_{cycle}$ reads 
\begin{eqnarray}
\tau_{cycle} &=& \tau_{1} + \tau_{2} + \tau_{1}^{R} + \tau_{2}^{R} \nonumber \\
&=& \frac{\kappa}{\omega_{1}} (1+\epsilon/\kappa) \left[ 1 + \left( \frac{\lambda_{1}}{\lambda_{2}}\right)^{1/2}\right]\,.
\end{eqnarray}

Thus, the minimum value of $\tau_{cycle}$ is attained when $\epsilon/\kappa\to 0$ since the relaxation times $\tau_{1,2}^{R}$ cannot be reduced without additional interference in the system. The power generated by the engine is then the work performed in the cycle divided by the cycle duration and reads (see for instance Ref.~\cite{abah2012,curzon1975})
\begin{eqnarray}
\mathcal{P}_{LR} &=& \frac{|W_{cycle}|}{\tau_{cycle}} \nonumber \\
&=& \frac{k_{B}T_{2} \omega_{1}}{\kappa} \frac{ \left[ 1-\left( \frac{\lambda_{1}}{\lambda_{2}}\right)^{1/2}\right] \left[1-\frac{T_{1}}{T_{2}}\left( \frac{\lambda_{2}}{\lambda_{1}}\right)^{1/2} \right]}{(1+\epsilon/\kappa)\left[ 1 + \left( \frac{\lambda_{1}}{\lambda_{2}}\right)^{1/2}\right]}\,, \nonumber \\
\label{eq:power}
\end{eqnarray}
which is maximum when $\epsilon/\kappa\to 0$ since Eq.~(\ref{eq:power}) would be given by the ratio between the largest possible value of work, namely, its value for a quasistatic cycle, and the least possible value of $\tau_{cycle}$.

\begin{figure}
\includegraphics[width=.41\textwidth]{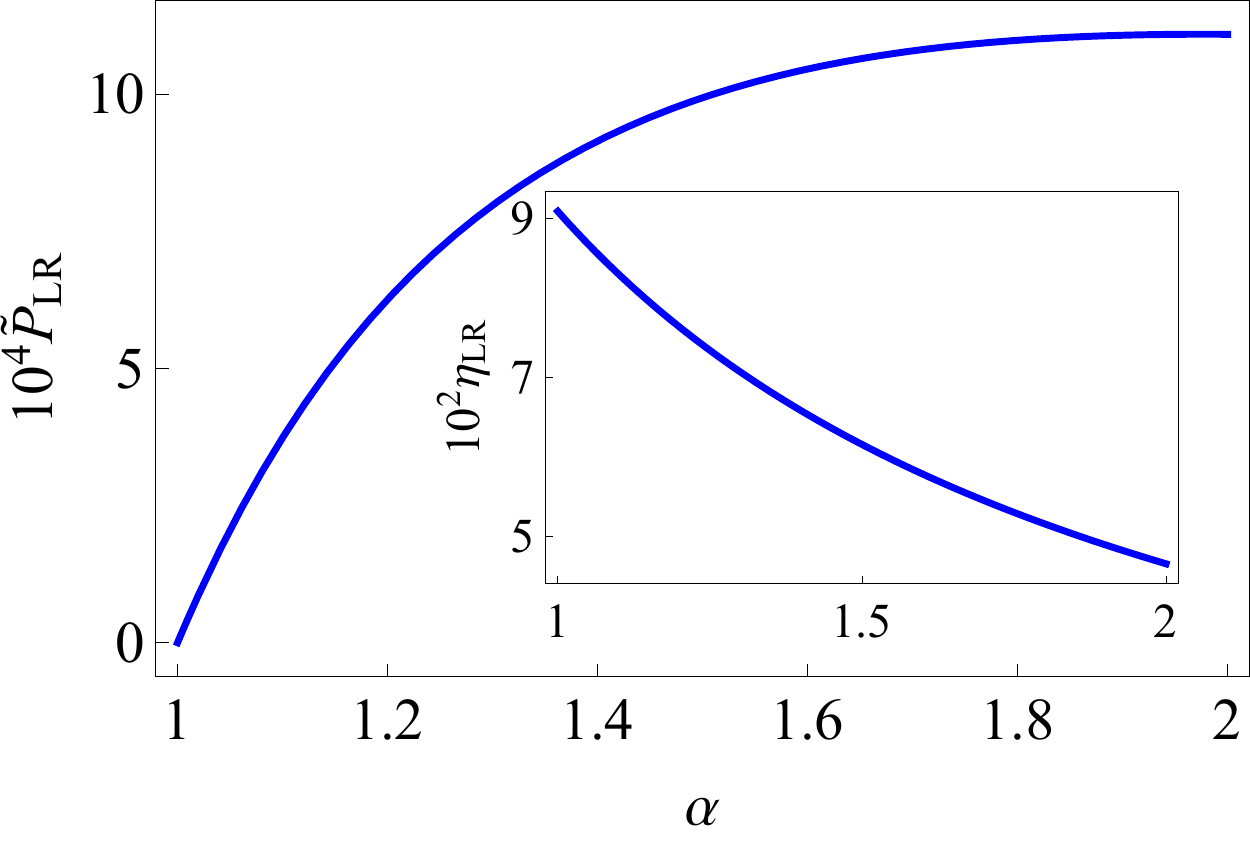}
\caption{\label{fig:power} (color online) Behavior of $\tilde{\mathcal{P}}_{LR} \equiv \kappa\mathcal{P}_{LR}/(k_{B}T_{2}\omega_{1})$, given by Eq.~(\ref{eq:power2}), and $\eta_{LR}=1-(T_{1}/T_{2})^{1/\alpha}$ as functions of $\alpha$ for $\delta T/T_{1} = 0.1$ when $\epsilon/\kappa\to 0$.}
\end{figure}

It is worth emphasizing at this point that the idea throughout our analysis is that the ratio $\lambda_{1}/\lambda_{2}$, or equivalently $\delta\lambda/\lambda_{1}$, is fixed (due to the linear response requirement $\delta\lambda/\lambda_{1}\ll 1$) and therefore Eq.~(\ref{eq:power}) cannot be optimized as a function of this parameter. The optimization procedure was already performed, for a fixed $\lambda_{1}/\lambda_{2}$ in the linear response regime, by choosing an appropriate $\lambda(t)$ that maximizes $W_{cycle}$ and at the same time minimizes $\tau_{cycle}$. However, we can optimize (\ref{eq:power}) as a function of $\alpha$. Using (\ref{eq:choice}), we can rewrite $\mathcal{P}_{LR}$ as 
\begin{equation}
\mathcal{P}_{LR} = \frac{k_{B}T_{2}\omega_{1}}{\kappa}\frac{ \left[ 1-\left( \frac{T_{1}}{T_{2}}\right)^{\frac{1}{\alpha}}\right] \left[1-\left( \frac{T_{1}}{T_{2}}\right)^{\frac{(\alpha-1)}{\alpha}} \right]}{(1+\epsilon/\kappa)\left[ 1 + \left( \frac{T_{1}}{T_{2}}\right)^{\frac{1}{\alpha}}\right]}\,,
\label{eq:power2}
\end{equation}
which already shows that power goes to zero as we approach Carnot efficiency, $\alpha\to 1$, if $k_{B}T_{2}\omega_{1}$ remains finite. This statement might suggest that Carnot efficiency at finite power is obtained by letting $k_{B}T_{2}\omega_{1}\to\infty$ as $\alpha$ approaches $1$. We are not sure that such indeterminacy leads to something conclusive. However, it is correct to say that the value of $k_{B}T_{2}\omega_{1}$ is not bounded and it can be chosen to provide finite values of maximum power for $\alpha$ very close to $1$. Figure~\ref{fig:power} illustrates the behavior of Eq.~(\ref{eq:power2}) and $\eta_{LR} = 1-(T_{1}/T_{2})^{1/\alpha}$ for a fixed $\delta T/T_{1}$ showing the trade-off between power and efficiency of our model.

%%%%%%%%%%%%%%%%%%%%%%%%%%%%%%%%%%%%%%%%%%%%%%%%%%%%%%%%%%%%%%%%%%%%%%%%%
%\section{Conclusions and Discussions}
\section{Conclusions}

We have constructed a heat engine under the following assumptions: (1) processes in which the working medium (a harmonic oscillator) is thermally isolated are performed at arbitrarily short times due to a specific kind of shortcuts to adiabaticity (without introducing extra energetic costs); (2) such processes simultaneously optimize the extracted work and time spent in the cycle; (3) the thermalization strokes, modeled as Brownian motion in the underdamped regime, essentially determine the cycle duration in terms of the corresponding relaxation times. The final result is a finite-time engine whose efficiency is identical to that of the quasistatic Otto cycle but at maximum power. By a suitable relation between the temperature and the compression ratios (see Eq.~(\ref{eq:choice})), this efficiency can be made arbitrarily close to Carnot's while keeping power at maximum. Although the whole cycle is formulated in linear response the efficiency breaks down previously derived bounds for this regime. The origin of this disagreement is the following: previous works have considered only a particular class of processes in the linear response regime, namely, those in which any delay between thermodynamic fluxes and forces is neglected. Once one starts with the most general formulation of linear response, given by Eq.~(\ref{eq:irrethermo}), effects due to both fast and slow processes can be described and new possibilities arise for efficiency at maximum power.

\begin{acknowledgments}
%\paragraph{Acknowledgments.}
The author acknowledges S. Deffner for several useful comments about the manuscript. The author also acknowledges support from FAPESP (Funda\c{c}\~ao de Amparo \`a Pesquisa do Estado de S\~ao Paulo) (Brazil) (Grant No. 2015/20194-0).
\end{acknowledgments}

%\appendix*

\appendix
\section{Linear-response shortcuts for the time-dependent harmonic oscillator}
\label{sec:app}

Inserting the relaxation function (\ref{eq:relaxf}) into the expression (\ref{eq:wexlr}) for the excess work, we obtain
\begin{eqnarray}
W_{ex} &\propto& \int_{t_{0,i}}^{t_{0,i}+\tau_{i}}dt\int_{t_{0,i}}^{t_{0,i}+\tau_{i}}dt'\,\cos[2\omega_{i}(t-t')]\,\dot{g}_{i}(t)\,\dot{g}_{i}(t')\nonumber\\
&=& \int_{0}^{1}ds \int_{0}^{1}ds'\,\cos[2\omega_{i}\tau_{i}(s-s')]\,g'_{i}(s)\,g'_{i}(s') \nonumber\\
&=& \left[ \int_{0}^{1}ds\,g'(s) \cos{(2\omega_{i}\tau_{i} s)}\right]^{2} + \left[ \int_{0}^{1}ds\,g'(s) \sin{(2\omega_{i}\tau_{i} s)}\right]^{2}\,.\nonumber\\
\label{eq.ap1}
\end{eqnarray} 
where we have defined $s = (t-t_{0,i})/\tau_{i}$, $s' = (t'-t_{0,i})/\tau_{i}$ and $g'(s)$ ($g'(s')$) denotes $dg/ds$ ($dg/ds'$).

This expression for $W_{ex}$ vanishes if each term of the sum in the last line of (\ref{eq.ap1}) is zero for the same value of $\tau_{i}$. To verify this possibility, we check under what conditions
\begin{equation}
z_{i} \equiv \int_{0}^{1}ds\,g'(s)\,e^{i2\omega_{i}\tau_{i} s} = 0\,.
\end{equation}

Using protocol (\ref{eq:protocg}), the quantity $z_{i}$ reads
\begin{eqnarray}
\lefteqn{z_{i} =} \nonumber \\
&&i 2\omega_{i}\tau_{i}\left\{ \frac{1-\cos(2\omega_{i}\tau_{i})}{(2\omega_{i}\tau_{i})^{2}} + 2\pi a_{i}\frac{[1-\cos(2\pi)\cos(2\omega_{i}\tau_{i})]}{(2\omega_{i}\tau_{i})^{2}-(2\pi)^{2}}\right\} \nonumber \\
&+& 2\omega_{i}\tau_{i} \left[ \frac{\sin(2\omega_{i}\tau_{i})}{(2\omega_{i}\tau_{i})^{2}} + 2\pi a_{i} \frac{\cos(2\pi)\sin(2\omega_{i}\tau_{i})}{(2\omega_{i}\tau_{i})^{2}-(2\pi)^{2}}\right].
\end{eqnarray}

Some zeros of $z_{i}$ appear whenever $2\omega_{i}\tau_{i} = 2\pi n$, with $n$ an integer greater than 1. These zeros are independent of the value of $a_{i}$. If we now demand that the real and imaginary parts of $z_{i}$ vanish for the \emph{same} value of $a_{i}$, we obtain Eq.~(\ref{eq:lrtau}). Hence, the excess work is zero for the arbitrarily short values of $\tau_{i}$ given by Eq.~(\ref{eq:lrtau}). In summary, the thermodynamic work performed along the finite-time protocols (\ref{eq:protocg}) may be identical to the quasistatic work.

%\bibliography{otto1bb}
%

\end{document}